\documentclass[aps,prl,reprint,twocolumn,preprintnumbers,superscriptaddress,letterpaper]{revtex4-2}
\usepackage[english]{babel}
\usepackage{ucs}
\usepackage[utf8x]{inputenc}
\usepackage{amsmath}
\usepackage{amsfonts}
\usepackage{amssymb}
\usepackage{graphicx}
\usepackage{bm}
\usepackage{bbm}
\usepackage{empheq}
\usepackage{xr-hyper}
\usepackage{hyperref}
\usepackage{xcolor}
\usepackage{chngcntr}

\makeatletter
\def\maketitle{
\@author@finish
\title@column\titleblock@produce
\suppressfloats[t]}
\makeatother

\newcommand{\ket}[1]{\left|#1\right\rangle}

\newcommand{\strategy}[1]{\mbox{\textsl{#1}}}

\newcommand{\goog}{\affiliation{Google Quantum AI, Santa Barbara, CA 93117, USA}}
\newcommand{\googla}{\affiliation{Google Quantum AI, Venice, CA 90291, USA}}
\newcommand{\ucsb}{\affiliation{Department of Physics, University of California, Santa
Barbara, CA 93106, USA}}
\newcommand{\ucr}{\affiliation{Department of Electrical and Computer Engineering, University of California, Riverside, CA 92521, USA}}
\newcommand{\umass}{\affiliation{{Department of Electrical and Computer Engineering, University of Massachusetts, Amherst, MA 01003, USA}}}
\newcommand{\auburn}{\affiliation{{Department of Electrical and Computer Engineering, Auburn University, Auburn, AL 36849, USA}}}

\renewcommand{\sec}[1]{\hyperref[sec:#1]{Section~\ref*{sec:#1}}}
\newcommand{\eqn}[1]{\hyperref[eqn:#1]{Equation~\ref*{eqn:#1}}}
\newcommand{\fig}[1]{\hyperref[fig:#1]{Figure~\ref*{fig:#1}}}
\newcommand{\tbl}[1]{\hyperref[tbl:#1]{Table~\ref*{tbl:#1}}}

\renewcommand{\thefigure}{\arabic{figure}}
\renewcommand{\theequation}{\arabic{equation}}
\renewcommand{\thetable}{\arabic{table}}
\renewcommand{\thesection}{\arabic{section}}

\newcommand{\fakepart}{
  \par\refstepcounter{part}
}

\begin{document}

\fakepart
\title{Overcoming leakage in scalable quantum error correction}
\let\oldauthor\author
\renewcommand{\author}[1]{\oldauthor{\mbox{#1}}}

\author{Kevin C.~Miao}
\thanks{These authors contributed equally to this work}
\goog

\author{Matt McEwen}
\thanks{These authors contributed equally to this work}
\goog
\ucsb

\author{Juan Atalaya}

\author{Dvir Kafri}
\author{Leonid P.~Pryadko}
\author{Andreas Bengtsson}
\author{Alex Opremcak}
\author{Kevin J.~Satzinger}
\author{Zijun Chen}
\author{Paul V.~Klimov}
\author{Chris Quintana}
\goog


% ...

\date{\today}

\begin{abstract}
Leakage of quantum information out of computational states into higher energy states represents a major challenge in the pursuit of quantum error correction (QEC). 
In a QEC circuit, leakage builds over time and spreads through multi-qubit interactions.
This leads to correlated errors that degrade the exponential suppression of logical error with scale, challenging the feasibility of QEC as a path towards fault-tolerant quantum computation. 
Here, we demonstrate the execution of a distance-3 surface code and distance-21 bit-flip code on a Sycamore quantum processor where leakage is removed from all qubits in each cycle. 
This shortens the lifetime of leakage and curtails its ability to spread and induce correlated errors.
We report a ten-fold reduction in steady-state leakage population on the data qubits encoding the logical state and an average leakage population of less than $1\times10^{-3}$ throughout the entire device. 
The leakage removal process itself efficiently returns leakage population back to the computational basis, and adding it to a code circuit prevents leakage from inducing correlated error across cycles, restoring a fundamental assumption of QEC. 
With this demonstration that leakage can be contained, we resolve a key challenge for practical QEC at scale.
\end{abstract}

\maketitle

Quantum error correction (QEC) promises to exponentially suppress uncorrelated errors in quantum computing devices, bridging the gap between achievable physical error rates and the low logical error rates required for useful quantum algorithms \cite{bravyi_quantum_1998, fowler_surface_2012, terhal_quantum_2015}. The surface code is a promising candidate for experimental implementations of QEC, where a repetitive stabilizer circuit protects a logical qubit state.

Superconducting transmon qubits \cite{koch_charge-insensitive_2007, kjaergaard_superconducting_2020} represent a leading platform for implementing surface code QEC, with recent demonstrations of architectures compatible with QEC and capable of scaling \cite{reed_realization_2012, chow_implementing_2014, corcoles_detecting_2014, riste_detecting_2015, kelly_state_2015,  takita_demonstration_2016, andersen_entanglement_2019, andersen_repeated_2020, google_quantum_ai_exponential_2021, sundaresan_matching_2022, krinner_realizing_2022, google_quantum_ai_suppressing_2022}.
However, the transmon is only weakly nonlinear, with transitions between successive states closely spaced in frequency. Transitions from the qubit \emph{computational} states to higher energy \emph{leakage} states are therefore difficult to avoid. These leakage states can be significantly populated by single-qubit gates \cite{motzoi_simple_2009, chen_measuring_2016}, entangling gates \cite{barends_superconducting_2014, neill_path_2017, yan_tunable_2018, rol_fast_2019, negirneac_high-fidelity_2021}, and measurement \cite{sank_measurement-induced_2016, shillito_dynamics_2022}.

Leakage is particularly dangerous in the context of QEC \cite{fowler_coping_2013, fowler_quantifying_2014, varbanov_leakage_2020, bultink_protecting_2020}. A key underlying assumption of QEC is that the physical errors to be suppressed are sufficiently uncorrelated in both space and time. Contrary to this requirement, a qubit in a leakage state can induce errors on multiple neighboring qubits, even causing them to leak as well \cite{varbanov_leakage_2020}. The correlated spread of errors through the device represents a major problem for experimental QEC. Identifying and post-selecting out leakage events has permitted cutting-edge experiments on the surface code \cite{krinner_realizing_2022, sundaresan_matching_2022}, and partial leakage removal has been integrated into surface code circuits \cite{google_quantum_ai_exponential_2021, google_quantum_ai_suppressing_2022}. However, all these experiments displayed a characteristic rise in the number of detected errors as the code progressed, indicative of accumulating leakage population in the device. A demonstration of leakage removal from all qubits in a surface code circuit has not yet been reported. Further, stabilizing the leakage populations such that error rates do not grow over time is a requirement for scalable QEC, and this remains an important open challenge.

Here, we study and remove the effects of leakage in a surface code circuit on an array of transmon qubits. First, we detail the dynamics of leakage in the QEC circuit and the spread of errors through space and time. We quantify the effect of leaked qubits undergoing multi-qubit interactions, which is the primary vehicle for spatial propagation of leakage.
Second, we demonstrate the effective removal of leakage from all qubits involved in the surface code circuit. We show residual leakage populations averaged over all qubits are suppressed to below $1\times10^{-3}$, and do not grow as the code is extended in time. 
Finally, we show that removing leakage improves logical performance. 
Using a distance-21 bit-flip code with leakage removal, injected leakage impacts logical performance equivalently to injected Pauli errors. This confirms that leakage removal is effective in suppressing the correlated nature of leakage-induced errors.
Then, using a distance-3 surface code, we show that leakage removal both decreases the rate of logical errors and prevents the code performance from declining over time, proving that QEC can be stable when carried out over many cycles. We extrapolate this behavior to larger code distances operating well below threshold, where we find that injected leakage impacts logical error rates in the same fashion as uncorrelated Pauli errors.
In summary, leakage removal overcomes an important obstacle to growing QEC to algorithmically relevant scales.

\section{Characterizing the spread of leakage}\label{sec:spread} 

\begin{figure}[t!]
    \centering
    \includegraphics[width=0.48\textwidth]{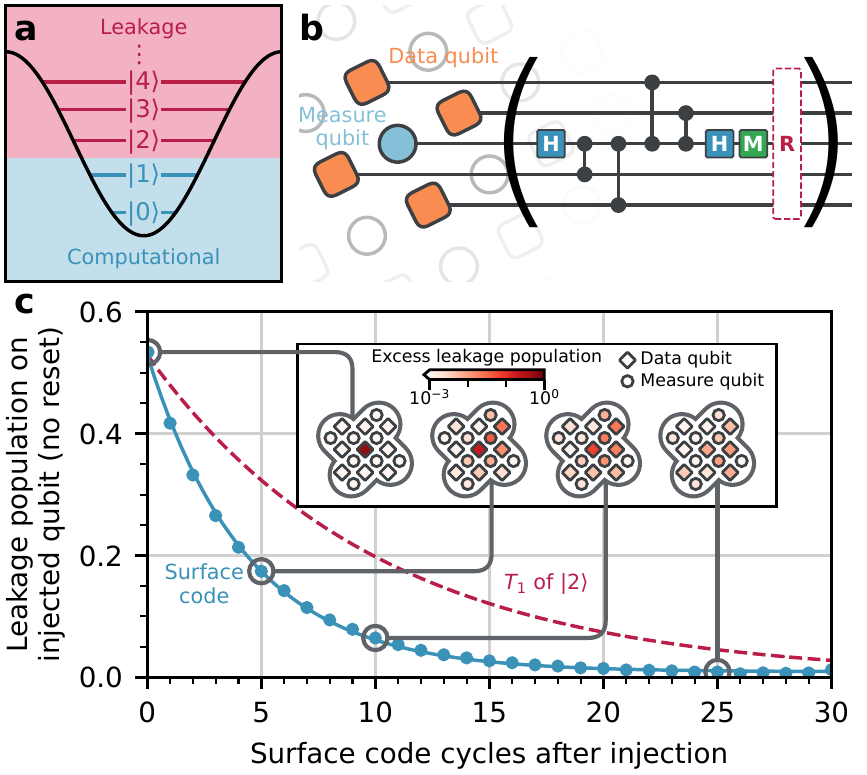}
    \caption[Leakage in a structured QEC circuit. ]{
    \textbf{Leakage in a structured QEC circuit. }
        a) The energy potential of a transmon qubit, illustrating the computational energy levels $\ket{0}$ and $\ket{1}$ (blue) and the leakage levels $\ket{2}$ and higher (red).
        b) The circuit for surface code QEC, showing a square grid comprised of measure qubits (light blue circle) and data qubits (orange squares). The cycle consists of four layers of entangling gates, along with intervening single qubit rotations, followed by the measurement (\textbf{M}) and reset (\textbf{R}). The reset operation here is shown across all qubits; it may be implemented as single qubit operations on the measure qubit, or include entangling operations with various neighboring data qubits.
        c) The time decay (main, blue) and spatial spread (inset) of leakage in a distance-3 surface code following the injection of $\ket{2}$ on the central data qubit. Each cycle takes approximately 1 $\mu$s, and leakage population is measured at the end of each cycle. The expected decay of $\ket{2}$ from $T_1$ relaxation on the leaked qubit alone is indicated (dashed red). 
        Excess leakage population is defined as the subtraction of leakage population in the absence of injection from the leakage population in the presence of injection.
    }
    \label{fig:leakage_spread}
\end{figure}

Leakage states (\fig{leakage_spread}a) are particularly problematic in structured QEC circuits because they are long-lived and spread through the device, inducing correlated errors in both space and time.
The surface code circuit displayed in \fig{leakage_spread}b shows a single cycle, which consists of a number of moments. A moment is a grouping of gates operated concurrently in time. Four such moments correspond to CZ gates used to measure the surface code stabilizers. When a qubit in the circuit leaks, subsequent gates involving that qubit produce additional errors. 

\fig{leakage_spread}c illustrates the dynamics of leakage in a distance-3 surface code circuit. At the cycle labeled $0$, we inject a full $\ket{1} \rightarrow \ket{2}$ rotation on the central data qubit, producing an expected near-50\% $\ket{2}$ population.
It takes many surface code cycles before this injected leakage population decays sufficiently, with an exponential decay constant around $4.4$ cycles. However, this decay is somewhat faster than the expected decay from $T_1$ relaxation of $\ket{2}$ alone. The insets show that the leakage population does not stay on the injected qubit, but is also transported to neighboring qubits as the circuit progresses. At the small code distance being considered, this transport is enough to affect every qubit involved in the circuit. 

Without any attempt to remove it, a single leakage event persists for many rounds and spreads a significant distance through the device, affecting many measurements and inducing many error detection events. The number of uncorrelated errors required to produce the same effect is the \emph{decomposed weight} of the leakage event \cite{fowler_coping_2013}. This high weight of leakage events when decomposed into uncorrelated errors makes them especially problematic for QEC.  

\begin{figure}[t!]
    \centering
    \includegraphics[width=0.48\textwidth]{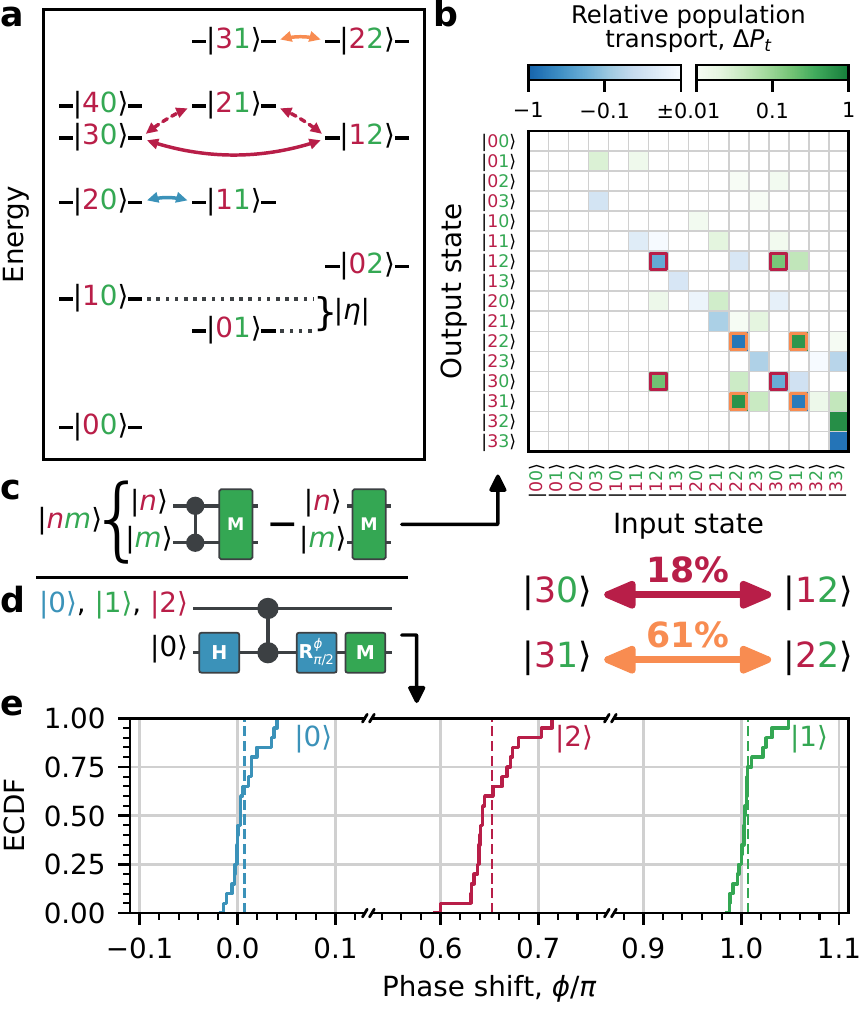}
    \caption[Leakage transport and phase errors in CZ gates. ]{
        \textbf{Leakage transport and phase errors in CZ gates. }
        a) The eigenenergy ladder for a pair of qubits satisfying the resonance condition for a diabatic CZ gate, where the qubits are detuned by their common nonlinearity $\left|\eta\right|$. We denote the two-qubit states $\ket{HL}$ with the higher (lower) energy qubit first (second). In addition to the intended resonance ($\ket{20}\leftrightarrow\ket{11}$), higher levels also satisfy a resonance condition, either directly ($\ket{31}\leftrightarrow\ket{22}$) or mediated by a two-photon process ($\ket{30}\leftrightarrow\ket{12}$).
        b) The relative population transport (net change in state populations) $\Delta P_t$ for the diabatic CZ gate, including the first two leakage levels. The rotation in $\ket{20}\leftrightarrow\ket{11}$ has been calibrated to $2\pi$. Highlighted are the off-diagonal elements due to the couplings between higher levels, with average relative population transport $\overline{|\Delta P_t|}$ shown below.
        c) The two circuits used to measure the relative population transport shown in (b). We subtract the population transport $P_t$ in the baseline experiment without a CZ gate (right) from the experiment with a CZ gate (left).
        d) The circuit for the modified Ramsey experiment shown in (e) with an interleaved CZ gate to a neighboring qubit at a higher frequency, followed by tomography on the lower frequency qubit.
        e) The measured phase shift $\phi$ during the modified Ramsey experiment with the neighboring qubit prepared in $\ket{0}$, $\ket{1}$, or $\ket{2}$, shown in an ECDF over 20 qubit pairs, with the mean value indicated by the dashed line. The CZ gate should produce a phase shift of $\phi = 0$ for an input $\ket{0}$, and a shift of $\phi = \pi$ for an input $\ket{1}$.
        A spurious phase shift near $\phi \approx 0.65\pi$ is produced when the higher-energy qubit is prepared in $\ket{2}$.
    }
    \label{fig:leakage_transport}
\end{figure}

The precise dynamics of leakage depends primarily on the details of the entangling gate used in the circuit. Here, we focus on the diabatic CZ gate used in the Sycamore architecture \cite{foxen_demonstrating_2020, google_quantum_ai_exponential_2021, google_quantum_ai_suppressing_2022}. This gate involves biasing qubits to satisfy the resonance conditions indicated in \fig{leakage_transport}a, and tuning the interaction strength to achieve a rotation of $2\pi$ in $\ket{11}\leftrightarrow\ket{20}$. We maintain the convention that the higher-energy qubit state is listed first in two-qubit states $\ket{HL}$. This resonance condition also aligns other resonances that involve leakage states. In particular, the $\ket{30}\leftrightarrow\ket{12}$ resonance enables a two-photon process which allows $\ket{2}$ on the lower-energy qubit to move to $\ket{3}$ on the higher-energy qubit. Similarly, the $\ket{31}\leftrightarrow\ket{22}$ resonance enables $\ket{3}$ on the higher-energy qubit to cause the lower-energy qubit to leak to $\ket{2}$, while the higher-energy qubit remains leaked in $\ket{2}$. These so-called \emph{leakage transport} processes are what allow leakage to spread, even in a single QEC cycle.  

The amount of leakage transport a gate produces is not normally calibrated, and so depends on the chosen gate length and effective coupling between levels. \fig{leakage_transport}b shows how a calibrated CZ gate affects populations, as measured by the circuits shown in \fig{leakage_transport}c.  In this device, we find around 18\% of the population of $\ket{30}$ is transported to $\ket{12}$ and vice versa.  The transport population is around 61\% for $\ket{31}\leftrightarrow\ket{22}$. We can also see the first indications of expected higher resonances such as $\ket{42}\leftrightarrow\ket{33}$. Data for each individual experiment and further characterisation of the readout can be found in Supplementary Information Section S1.

Even in the absence of leakage transport, we find that leakage induces additional errors in the CZ gate. When the higher-energy qubit is in $\ket{2}$ and the lower-energy qubit is in the computational basis, leakage transport is not possible but a significant phase error is imparted on the non-leaked qubit. When a CZ gate is applied as in \fig{leakage_transport}d with the higher-energy qubit in $\ket{0}$, we expect to see no phase shift $\phi = 0$ on the lower-energy qubit. With the higher-energy qubit prepared in $\ket{1}$, we expect to see a phase shift $\phi = \pi$, indicating a well-calibrated CZ gate. \fig{leakage_transport}e shows the relative phase for 20 pairs of qubits. When computational states $\ket{0}$ and $\ket{1}$ are prepared we see tight groupings around the expected phase shifts $\phi = 0$ and $\phi = \pi$, respectively. However, when a leakage state is prepared on the higher-energy qubit, we see a phase shift near $\phi \approx 0.65\pi$. This represents a significant computational error on the non-leaked qubit, and is a significant source of errors to be detected and corrected as leakage spreads. 

These results illuminate the dangers of leakage: A single leakage event on any qubit will expose many CZ gates to a leaked input state before it decays sufficiently. Each of these interactions has a significant probability to introduce new computational errors, move the leakage to another qubit, or induce additional leakage on previously non-leaked qubits. In QEC circuits, these effects are damaging enough that they must be included in simulations to achieve good agreement with experimental performance \cite{google_quantum_ai_suppressing_2022}. Accordingly, we are motivated to remove leakage in the code circuit so as to suppress these effects.

\section{Suppressing leakage populations during a QEC circuit}\label{sec:suppression}

Having better understood the dangers of leakage in QEC circuits, we turn to removing it. 
An unconditional reset gate can remove all energy from a qubit, including when it starts in a leakage state,  
and can be applied to the measure qubits at the end of each cycle \cite{reed_fast_2010, geerlings_demonstrating_2013, magnard_fast_2018, mcewen_removing_2021, zhou_rapid_2021}. However, our study of leakage transport motivates the need to remove leakage from the data qubits as well. Leaving the computational state intact is incompatible with unconditional reset and requires a more delicate \emph{leakage removal} operation.

Three broad approaches for leakage removal have been proposed: swap-type \cite{fowler_coping_2013, brown_leakage_2019}, where the roles of measure and data qubits are exchanged at a regular interval by the use of additional operations; feedback-type \cite{varbanov_leakage_2020, bultink_protecting_2020} where the leakage is identified classically from measurement patterns and feedback is applied to return the qubit to the computational subspace; 
and direct-type \cite{battistel_hardware-efficient_2021} where an operation is used to remove leakage from a qubit without disturbing the computational states.
In light of our findings on leakage transport, swap-type strategies become more difficult to justify; only half the qubits are reset in each cycle, and so leakage may still move between qubits and thereby spread through time.
Similarly, the conditional nature of feedback-type approaches prevents them from fully solving the leakage problem -- leakage states cause several errors before they are noticed and corrected.
Hence, we pursue a direct removal approach.

In the following sections, we present and compare three leakage removal strategies. First, \strategy{No Reset} forgoes any operations at the end of the cycle, representing the best case for a simple Pauli error model, but the worst case for leakage. Second, \strategy{MLR} applies multi-level reset (MLR) gates \cite{mcewen_removing_2021} on measure qubits immediately after measurement at the end of every cycle. This adds additional error to the cycle due to additional data qubit idle time while the gate is performed, but has been previously shown to remove leakage population and improve logical performance compared with the baseline \strategy{No Reset} strategy \cite{mcewen_removing_2021}. Finally, in \strategy{DQLR} we perform a multi-level reset on the measure qubits followed by data qubit leakage removal (DQLR), consisting of a two-qubit interaction to transport leakage from data to measure qubits and a reset gate on measure qubits. Additional details on the DQLR process and constituent operations are included in Supplementary Information Section S2.

\begin{figure}[t!]
    \centering
    \includegraphics[width=0.48\textwidth]{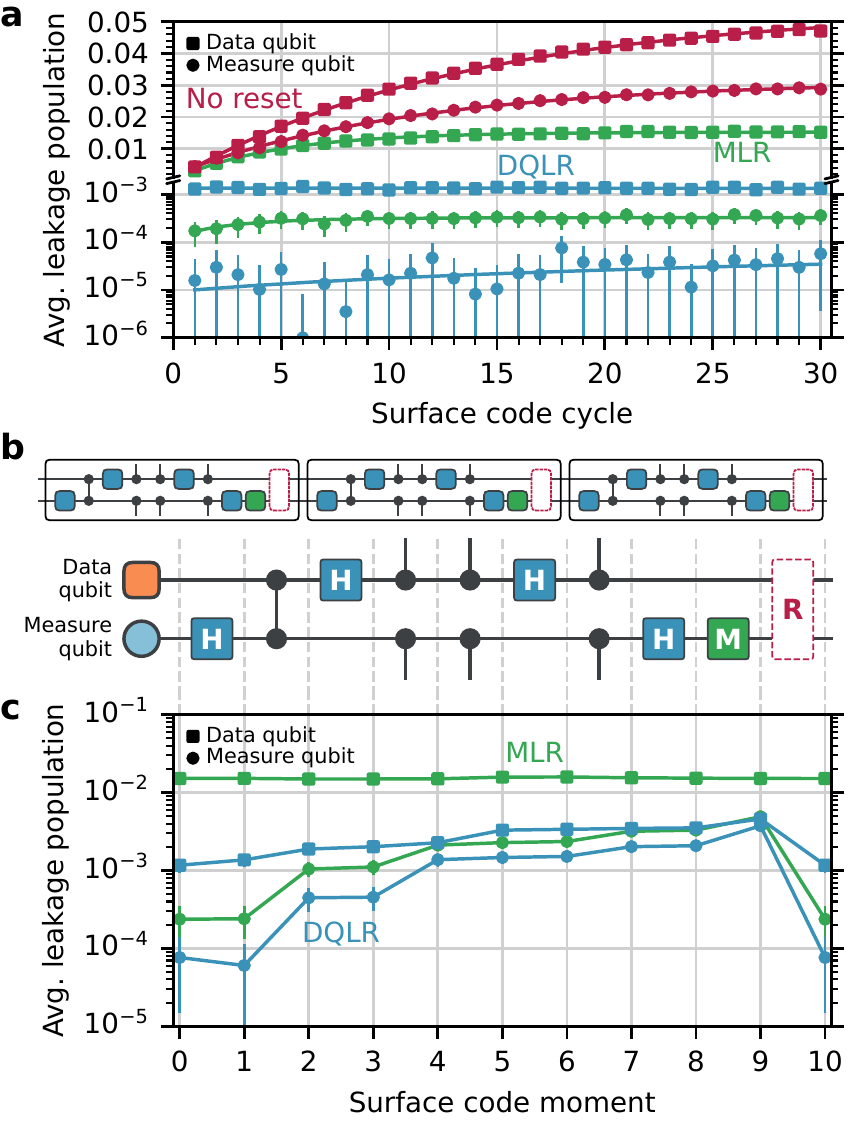}
    \caption[Leakage population during surface code execution. ]{
        \textbf{Leakage population during surface code execution. }
        a) Average leakage populations for data qubits (squares) and measure qubits (circles) measured at the end of each surface code cycle with \strategy{No reset} (red), \strategy{MLR} (green), and \strategy{DQLR} (blue). 
        b) Top: The surface code circuit shown for a pair of neighboring measure and data qubits. Each surface code cycle is highlighted (rounded rectangles). Bottom: A single surface code cycle showing each moment in the cycle.
        c) Leakage populations after each moment in the cycle for \strategy{MLR} (green) and \strategy{DQLR} (blue) leakage removal strategies, averaged over data qubits (squares) and measure qubits (circles) and over cycles 25--30.
    }
    \label{fig:leakage_slicing}
\end{figure}

To compare the leakage dynamics for the three strategies, we implement a \mbox{distance-3} surface code on a Sycamore processor. We measure the evolution of leakage population as the surface code progresses by truncating the circuit in time and performing a measurement that can resolve $\ket{2}$ on all qubits \cite{mcewen_removing_2021}. In \fig{leakage_slicing}a, we perform this truncation at the end of each surface code cycle (top of \fig{leakage_slicing}b). Using \strategy{No Reset}, we observe a gradual rise in leakage populations over all qubits, reaching nearly 5\% average leakage population for data qubits and nearly 3\% for measure qubits over 30 cycles. We note that, even after 30 cycles, leakage populations have not stabilized and continue to grow. Using \strategy{MLR} reduces average measure qubit leakage populations to about $3 \times 10^{-4}$, but average data qubit populations still rise to over 1.5\%. Using \strategy{DQLR} suppresses average leakage populations to around $10^{-3}$ for data qubits and less than $10^{-4}$ for measure qubits. Most importantly, \strategy{DQLR} maintains these levels throughout the full 30 cycles.

We can use the same technique to study the dynamics of leakage within a surface code cycle, by truncating the circuit at each moment midway through a cycle (bottom of \fig{leakage_slicing}b). \fig{leakage_slicing}c shows the leakage population measured after each moment in the cycle, averaged over cycles 25--30 where the leakage populations have stabilized. We neglect the \strategy{No Reset} strategy here, as leakage populations do not stabilize. With \strategy{MLR}, the average leakage population on the data qubits saturates to a stable value around 1.5\%, consistent with \fig{leakage_slicing}a. However, the average measure qubit leakage population starts each cycle at a very low value near $2\times10^{-4}$, grows over the course of the cycle as operations produce leakage, and is then reduced back to its initial low value by the reset procedure. This lets us estimate that the operations produce around $5\times10^{-3}$ leakage in each cycle. With \strategy{DQLR}, we see that leakage populations for both measure and data qubits grow over the course of the cycle, and are removed by the reset procedure. The data qubits start each cycle with around $1\times10^{-3}$ leakage population, again increasing to around $5\times10^{-3}$ immediately following measurement, before it is removed. The measure qubits attain even lower leakage populations compared to \strategy{MLR}.

These results demonstrate that our DQLR procedure successfully suppresses steady-state leakage populations to previously unachievable levels and stabilizes those levels over the course of a long QEC circuit. The removal strategy also contains the leakage dynamics to a single cycle. However, the residual ability for leakage to spread and 
induce correlated errors within a single round \cite{fowler_quantifying_2014} should be the subject of further study. 

\section{Effect on QEC logical performance}\label{sec:qec}

Having achieved low leakage populations in both data qubits and measure qubits with our DQLR procedure, we turn to evaluating logical performance. We consider two codes providing complementary information: a distance-21 bit-flip code and a distance-3 surface code. Our physical qubit error rates place the surface code close to threshold, whereas the bit-flip code is well below threshold \cite{google_quantum_ai_exponential_2021, google_quantum_ai_suppressing_2022}. The vastly lower logical error rates for the bit-flip code give us finer resolution on the effect of leakage within the code. In contrast, the surface code is a more challenging circuit for calibration and operation, and is sensitive to both bit-flip and phase-flip errors, providing an environment where more potentially adverse effects from reset can be detected and measured.

\begin{figure}[t]
    \centering
    \includegraphics[width=0.48\textwidth]{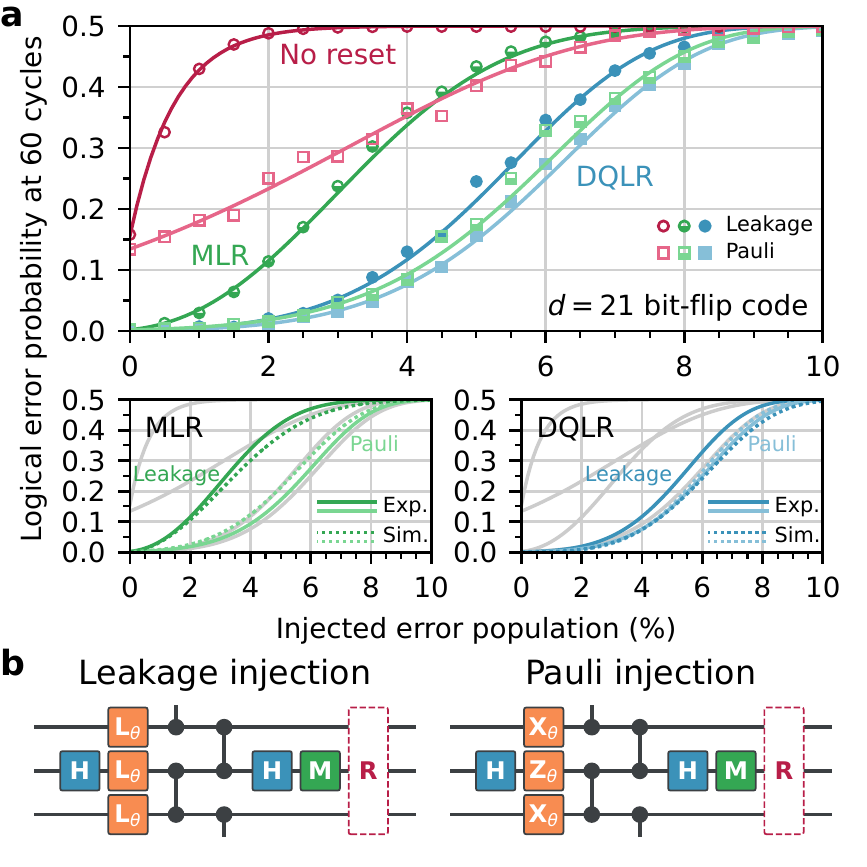}
    \caption[Bit-flip code logical performance and dependence on injected errors. ]{
        \textbf{Bit-flip code logical performance and dependence on injected errors. }
        a) Logical error probability for a distance-21 bit-flip code run for 60 cycles, under the effect of either injected leakage (dark circles) or injected Pauli errors (light squares). Three leakage removal strategies, \strategy{No reset} (red, unfilled), \strategy{MLR} (green, semi-filled), and \strategy{DQLR} (blue, filled), are considered. Lines are fits to experimental data using a power law with an offset.
        Below: Highlights of fits to experimental data (solid) and numerical simulations (dashed) for the \strategy{MLR} and \strategy{DQLR} strategies.
        b) Circuits for the bit-flip code, showing the error injection locations for both leakage (left) and Pauli errors (right).
    }
    \label{fig:qec_performance_bf}
\end{figure}

\fig{qec_performance_bf}a shows the logical error probability of a distance-21 bit-flip code carried out to 60 cycles while introducing both leakage and Pauli errors. 
We inject leakage population $P_L$ into all qubits by applying a $\ket{1} \leftrightarrow \ket{2}$ rotation on each qubit immediately after the first Hadamard gate layer (\fig{qec_performance_bf}b, left), where the rotation angle $\theta_L$ is
\begin{align*}
    \theta_L &= 2\sin^{-1}\left(\sqrt{2P_L}\right).
\end{align*}
We compare $P_L$ to injected Pauli error ``population'' $P_P$, which is produced by $X$ and $Z$ rotations on the data and measure qubits (\fig{qec_performance_bf}b, right), respectively, taking advantage of the classical nature of the bit-flip code. The Pauli error rotation angle $\theta_P$ is
\begin{align*}
    \theta_P &= 2\sin^{-1}\left(\sqrt{P_P}\right),
\end{align*}
where the missing factor of 2 relative to the definition of leakage population accounts for Pauli rotations always affecting the qubit state in the computational basis, whereas leakage injection only applies to qubit population in $\ket{1}$.
We fit the experimental data and numerical simulations to an offset power law as a guide, as detailed in Section S5 of the Supplementary Information.

With \strategy{No Reset}, even small amounts of injected leakage population less than 1\% cause the logical error probability to rise above 40\%. This is in contrast with correctable Pauli errors, which can be introduced to around 5\% population before similar logical error probabilities are encountered. With \strategy{MLR}, the logical error probability is drastically lowered without injection, consistent with prior measurements in bit-flip codes \cite{mcewen_removing_2021}. Still, the logical error probability rises much more rapidly when injecting leakage compared to injecting Pauli errors. We attribute this to unmitigated leakage accumulation on the data qubits, which leads to high decomposed weight of uncorrelated errors and ultimately logical errors. When we prevent this leakage buildup with \strategy{DQLR}, we observe a much smaller difference between the code's response to injected leakage compared to injected Pauli errors. This is strong evidence that the DQLR operation has successfully reduced the decomposition weight of a leakage event to near 1. In this situation, leakage has around the same influence on logical performance as an equivalent amount of Pauli error, and has been prevented from effectively spreading and inducing correlated errors.

We also note the good agreement between data and numerical simulation for injected leakage and Pauli errors, quantifying our understanding of the effects of leakage in the code with both \strategy{MLR} and \strategy{DQLR} strategies. In both cases, we note that we slightly underestimate the logical error induced by injected leakage, illustrating the difficulties of fully capturing the effect of correlated errors even with \strategy{DQLR} preventing substantial spread across cycles, and emphasising the importance of future work on leakage dynamics inside a single cycle. Nonetheless, the close correspondence of the Pauli simulation to the injected leakage experimental data for \strategy{DQLR} helps justify future Pauli simulations as useful estimates of final code performance when leakage is removed each cycle.

\begin{figure}[t]
    \centering
    \includegraphics[width=0.48\textwidth]{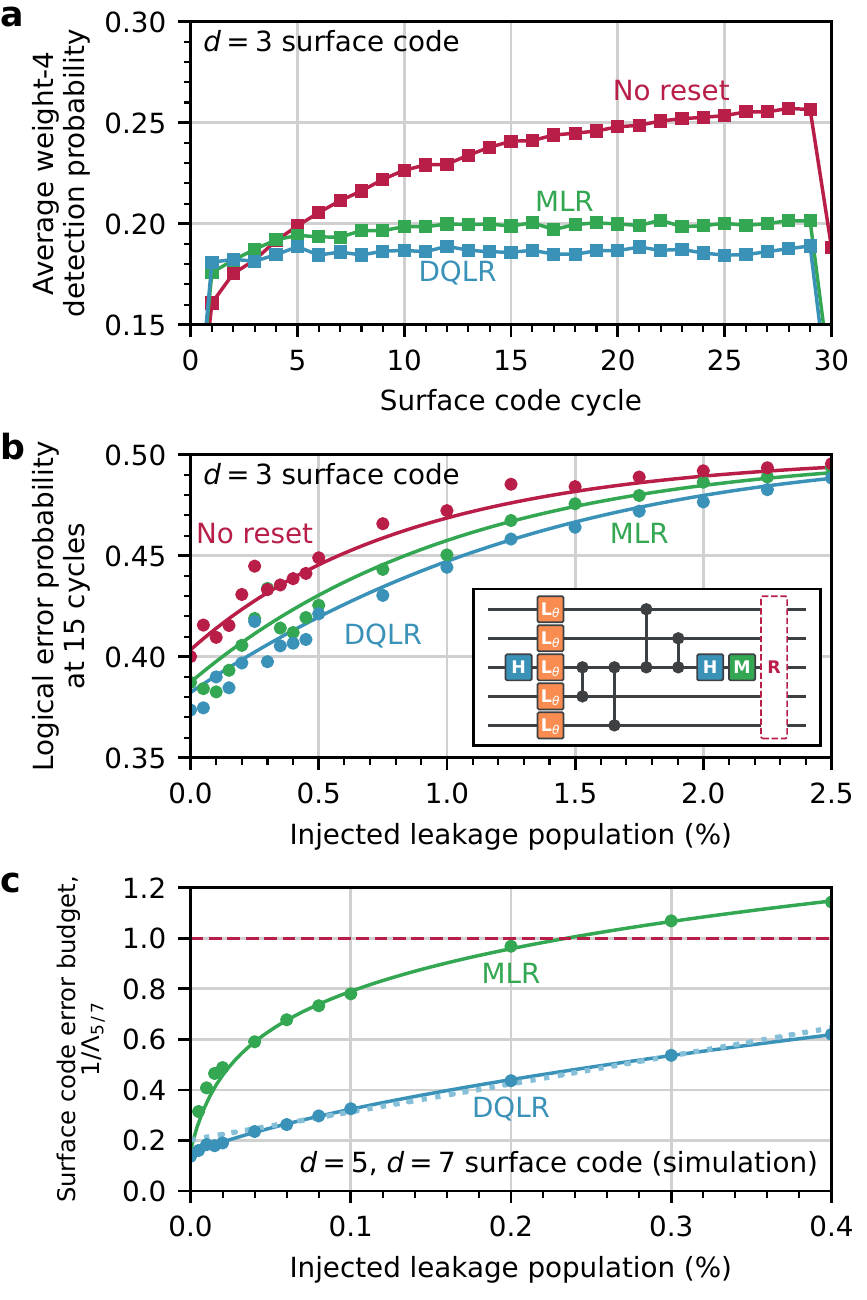}
    \caption[Surface code logical performance and dependence on injected errors. ]{
        \textbf{Surface code logical performance and dependence on injected errors. }
        a) The detection probability averaged for the weight-4 stabilizers in a distance-3 surface code under the three leakage removal strategies studied in this work.
        b) Logical error probability for a distance-3 surface code run for 15 cycles, under varying injected leakage population and the three different leakage removal strategies studied in this work. The inset shows that the circuit has an included layer where leakage is injected by performing a $\ket{1}\leftrightarrow\ket{2}$ rotation.
        c) Dependence of projected surface code error budget $1/\Lambda_{5/7}$ (the inverse of the exponential error suppression factor between a distance-5 and distance-7 surface code) under injected leakage population compared between \strategy{MLR} (green) and \strategy{DQLR} (blue). Solid lines are fits to a ratio of offset power laws, while the dotted light blue line is a linear fit of the data using \strategy{DQLR}.
    }
    \label{fig:qec_performance_sc}
\end{figure}

\fig{qec_performance_sc}a shows the average detection probabilities corresponding to the weight-4 stabilizers in the distance-3 surface code. Detection probabilities are the fraction of the total number of experiments where an error was detected on a given stabilizer. With \strategy{No Reset}, the buildup of leakage population produces more errors as the code progresses, creating a rising pattern of detection probability. With \strategy{MLR}, a large portion of this rise is mitigated, but the detection probability still rises by 2.5\% over the course of the first 15 cycles. With \strategy{DQLR} the detection probability immediately stabilizes to around 18\% and remains steady throughout the code duration. We attribute this to the recurrent removal of leakage on all qubits preventing growth in leakage populations and resulting correlated errors over time. This resolves a key concern in state of the art QEC work \cite{krinner_realizing_2022, sundaresan_matching_2022, google_quantum_ai_suppressing_2022} where detection probabilities were found to rise even with partial leakage removal or post-selection. These results confirm the relationship between rising detection probability and rising leakage populations and demonstrate the resolution of this effect.

In \fig{qec_performance_sc}b, we evaluate the three leakage removal strategies by measuring the logical error probability of a distance-3 surface code after 15 cycles.
At 0\% injected leakage the circuit corresponds to the standard code circuit with an additional idle where the injection is otherwise inserted. Over the range of injected leakage population values, \strategy{No Reset} exhibits the worst logical performance, followed by \strategy{MLR}, with \strategy{DQLR} having the lowest logical error probability. This confirms that \strategy{DQLR} improves logical errors by suppressing correlated errors from leakage, despite the additional cycle time and errors introduced by the DQLR operations. Further, \strategy{No Reset} and \strategy{MLR} degrade in logical performance faster with more injected leakage when compared to \strategy{DQLR}.

In order to study surface code performance in a regime further below threshold, we turn to numerical simulations of distance-5 and distance-7 surface codes. To consider scaling performance, we use the exponential error suppression factor $\Lambda_{5/7}$, defined as $\Lambda_{5/7} = \varepsilon_5 / \varepsilon_7$,
where $\varepsilon_5$ and $\varepsilon_7$ are the logical error rates for a distance-5 and distance-7 surface code, respectively. In \fig{qec_performance_sc}c, we investigate $\Lambda_{5/7}$ for a hypothetical device with lower component errors than what is currently realizable (see Supplementary Information Section S6 for details). In particular, we set intrinsic leakage rates to zero and vary the probability of leakage injection. With no leakage in the system, $\Lambda_{5/7} \approx 7.2$, independent of leakage removal strategy. However, when injecting up to $4\times10^{-3}$ leakage population per round (comparable to intrinsic leakage rates in current devices), the surface code error budget $1/\Lambda_{5/7}$ \cite{google_quantum_ai_suppressing_2022} rises rapidly and nonlinearly for \strategy{MLR}. In contrast, with \strategy{DQLR}, leakage increases $1/\Lambda_{5/7}$ much more slowly and with a near-linear dependence on injected leakage population, characteristic of an uncorrelated error source~\cite{google_quantum_ai_exponential_2021, google_quantum_ai_suppressing_2022}. With this ability to maintain effective error suppression in the presence of leakage, \strategy{DQLR} successfully mitigates the dangers of correlated leakage-induced errors to scalable QEC.

\section{Summary and Outlook}

We have demonstrated the effective removal of leakage from all qubits involved in a surface code QEC circuit. Moreover, we have shown that when leakage is removed on all qubits, correlated leakage-induced errors are suppressed. At the same time, the logical performance of the code improves outright and stabilizes in time. We confirm the conjecture that growth in logical errors is attributable to leakage, and we do not uncover other major sources of logical error that grow as the code continues in time. 

With these findings, we unequivocally resolve the longstanding concern that qubits with weak nonlinearity cannot successfully implement QEC at long times due to correlated leakage-induced errors.
As such, we confirm that large arrays of transmon qubits are a viable and promising architecture for QEC at scale.

\section*{Declarations}

\subsection*{Acknowledgments}
The authors would like to acknowledge and thank Rami Barends for his early work and guidance on the subject of data qubit leakage removal. 

\subsection*{Competing interests}
The authors declare no competing interests.

\subsection*{Availability of data and materials}
The data necessary to reproduce the findings for this study are available upon reasonable request, or at \href{https://doi.org/10.5281/zenodo.7302032}{10.5281/zenodo.7302032}.

\subsection*{Author contributions}
K.C.M.\ and M.M.\ developed leakage removal operations, designed and performed the experiments, analyzed the data and prepared the manuscript.
J.A.\ provided theoretical models and analysis.
D.K.\ and L.P.\ provided simulation tools.
A.B., A.O., K.J.S., Z.C., P.V.K., C.Q.\ provided and maintained key components of the experimental setup.
D.S.\ and Y.C.\ provided technical guidance and helped prepare the manuscript.
All authors contributed to the fabrication process, experimental set-up and operation, and revision of the manuscript.
\bibliography{arXiv}
\clearpage

\counterwithin{figure}{part}
\counterwithin{section}{part}

\fakepart
\title{Supplementary information for \emph{Overcoming leakage in scalable quantum error correction}}
\maketitle
\renewcommand{\thefigure}{S\arabic{figure}}
\renewcommand{\theequation}{S\arabic{equation}}
\renewcommand{\thetable}{S\arabic{table}}
\renewcommand{\thesection}{S\arabic{section}}

\title{Supplementary information for \emph{Overcoming leakage in scalable quantum error correction}}

\section{Effects of leakage on diabatic CZ gate}\label{sup:CZ}

During the diabatic CZ gate, additional levels are placed on resonance and contribute to the leakage transport phenomenon depicted in the main text Figure 2. 
The resonance $\ket{30}\leftrightarrow\ket{12}$ allows a two-photon transition mediated by $\ket{21}$, which is detuned by around the nonlinearity $\eta$. If $g$ is the induced coupling between $\ket{11}$ and $\ket{20}$, then the effective couplings are:
\begin{align*}
    g_{\ket{30}\leftrightarrow\ket{21}} &= \sqrt{3}g, \\
    g_{\ket{21}\leftrightarrow\ket{12}} &= 2g, \\
    g_{\mathrm{eff}} = g_{\ket{30}\leftrightarrow\ket{12}} &= -g_{\ket{21}\leftrightarrow\ket{12}} \times g_{\ket{30}\leftrightarrow\ket{21}} / \eta.
\end{align*}
Then, for a CZ gate where $g$ is maintained for time $t$, the population transport $P_t$ for $\ket{30}\leftrightarrow\ket{12}$ can be estimated as
\begin{align*}
    P_t = \sin^2\left(g_{\mathrm{eff}}t\right).
\end{align*}

\begin{figure}[t]
    \centering
    \includegraphics[width=0.48\textwidth]{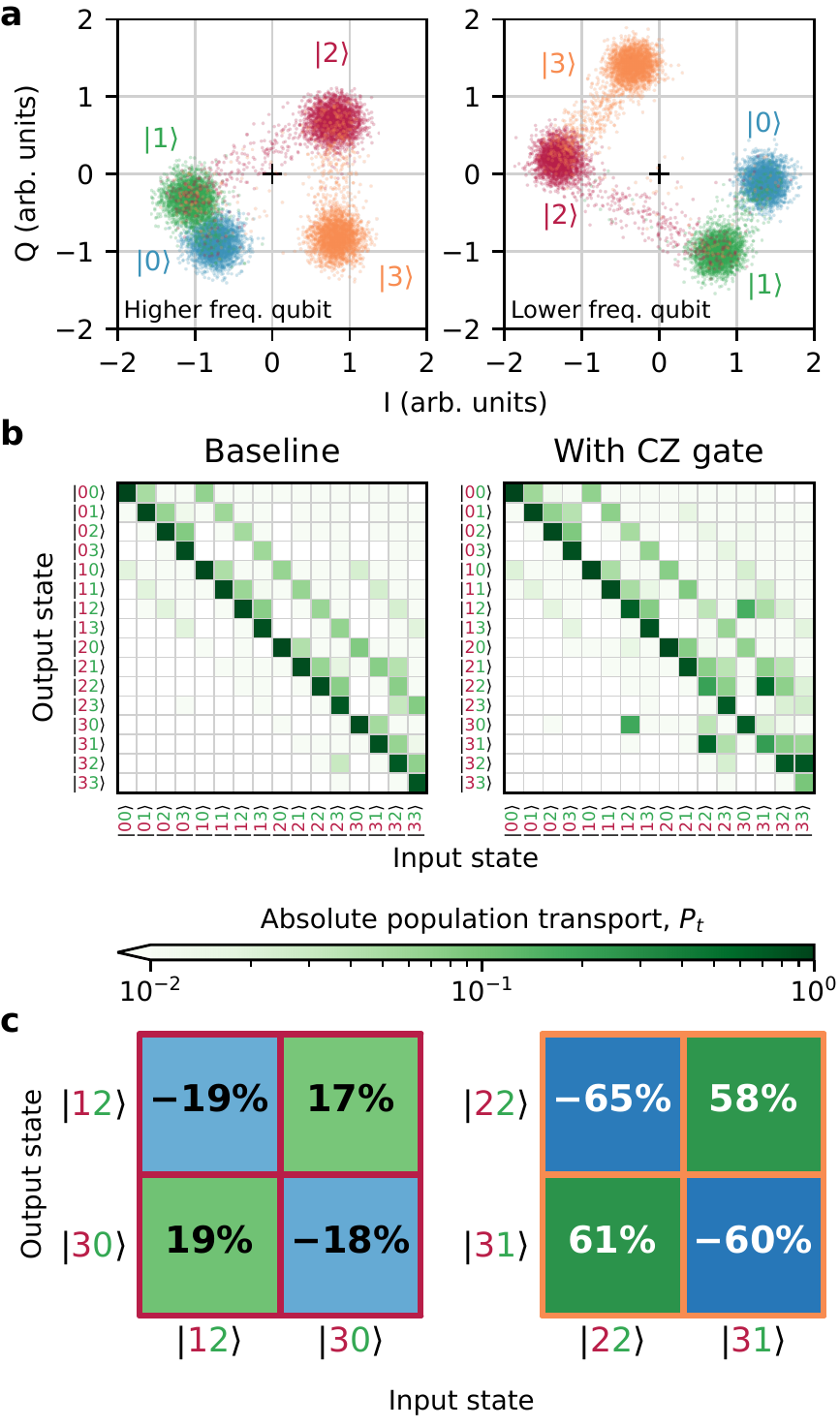} 
    \caption[Measuring leakage transport. ]{
    \textbf{Measuring leakage transport. }
        a) Readout clouds for the pair of qubits used in the leakage transport experiment, showing good distinguishability between all of the lowest four qubit energy states. 
        b) Raw measured population transport matrices for the two transport experiments. In both matrices, we can see the effect of $T_1$ decay during measurement, which is enhanced for the higher levels.
        Subtracting ``Baseline'' from ``With CZ gate'' produces the matrix shown in Figure 2b of the main text.
        c) Individual relative population transport values $\Delta P_t$ for relevant leakage transport mechanisms $\ket{30}\leftrightarrow\ket{12}$ and $\ket{31}\leftrightarrow\ket{22}$. The average of the norm of these values produce the population transport values at the bottom of Figure 2b of the main text.
    }
    \label{fig:leakage_transport_supp}
\end{figure}

To measure the leakage transport in the diabatic CZ gate, we calibrate a readout pulse capable of distinguishing all of the four lowest qubit energy 
levels, as shown in \fig{leakage_transport_supp}a. When we encounter $\ket{4}$ during measurement using this readout pulse, we assign and identify it as $\ket{3}$.
The baseline experiment consists of preparing a given two-qubit state using microwave drives and then performing simultaneous readout of the qubits. The ``Baseline'' matrix of \fig{leakage_transport_supp}b shows the results of this experiment, illustrating that performing readout simultaneously does not impact the high distinguishability between all two-qubit states. We can also see that the majority of the error in this simultaneous readout is due to $T_1$ decay during the readout process. These decay channels reduce the populations on the main diagonal by a few percent, and become more prominent for higher levels. The ``With CZ gate'' matrix of \fig{leakage_transport_supp}b shows the same experiment with a CZ gate inserted between state preparation and measurement. We then see new off-diagonal processes corresponding to leakage transport. We subtract the ``Baseline'' matrix from the ``With CZ gate'' matrix to produce the matrix shown in Figure 2b of the main text. The leakage transport processes that are most relevant to the spread of leakage in our system correspond to the $\ket{30}\leftrightarrow\ket{12}$ and $\ket{31}\leftrightarrow\ket{22}$ rotations. In \fig{leakage_transport_supp}c, we numerically show the individual relative population transport values $\Delta P_t$ for these two processes. We take the mean of the absolute value of the relative population transport values to determine the values at the bottom of Figure 2b of the main text.

\begin{figure}[t]
    \centering
    \includegraphics[width=0.48\textwidth]{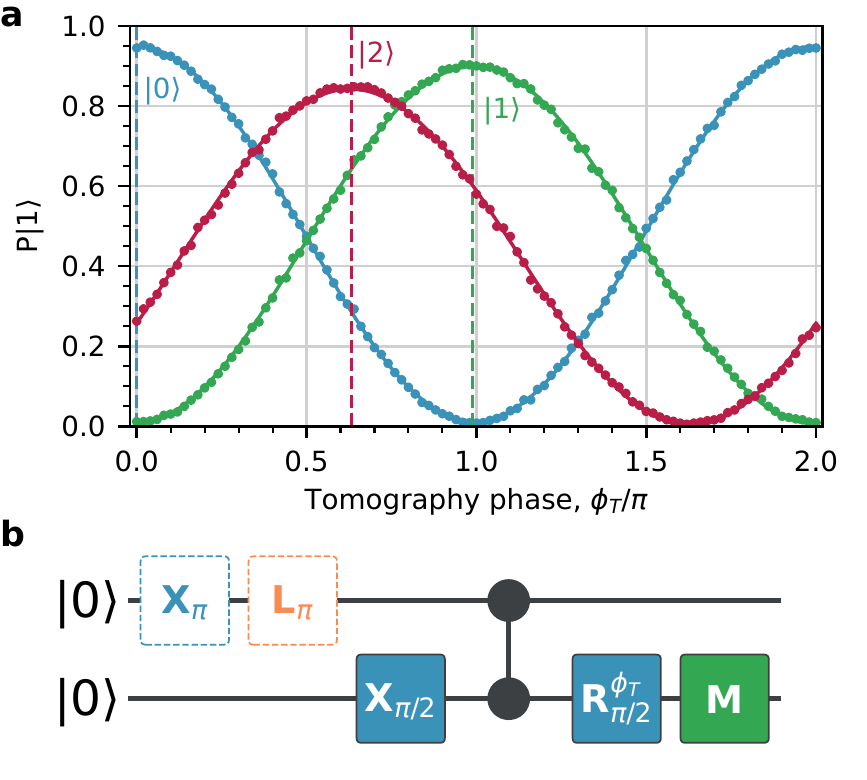}
    \caption[Measuring leakage phases. ]{
    \textbf{Measuring leakage phases. }
        a) Raw data for the leakage phase experiment on a single pair of qubits. Each of $\ket{0}$, $\ket{1}$, and $\ket{2}$ is prepared on the higher-energy qubit, and then a Ramsey experiment with an interleaved CZ gate is performed on the lower-energy qubit. The solid lines are sinusoidal fits and the dashed lines indicate the extracted phase shifts $\phi$ shown in the main text.
        b) The hardware circuit executed in the above experiment in the $\ket{2}$ case. The two initial rotations (\textbf{X} and \textbf{L}) on the higher-energy qubit (top) are removed when preparing $\ket{0}$, and the second rotation \textbf{L} is removed when preparing $\ket{1}$.
    }
    \label{fig:leakage_phase}
\end{figure}

To measure the spurious phase shift generated by leaked higher-energy qubits undergoing the CZ gate, we perform the experiment detailed in \fig{leakage_phase}a. The higher-energy qubit in the pair is prepared in each of $\ket{0}$, $\ket{1}$, and $\ket{2}$, while a Ramsey experiment is performed on the lower-energy qubit with an interleaved CZ gate. On the lower-energy qubit, we vary the tomography phase $\phi_T$ of the second pulse relative to the first pulse in the Ramsey experiment to record data as shown in \fig{leakage_phase}b for each input state. We fit a sinusoid to the experimental data and extract the phase offset $\phi$. We perform this experiment on 20 pairs of qubits, and show the empirical cumulative distribution function of the extracted phase offsets in Figure 2e of the main text.

\section{Leakage Removal Strategy Details}\label{sup:strategy}

We studied three leakage removal strategies; \strategy{No reset}, \strategy{MLR} and \strategy{DQLR}. We now describe them in greater detail.

For \strategy{No reset}, we add no additional operations at the end of each cycle. Because this prepares the qubit for the next cycle in whichever state was measured rather than deterministically in $\ket{0}$, this also requires the redefinition of the detectors in the surface and bit-flip code circuits: rather than comparing time-neighboring measurements, we compare time-next-neighboring measurements on the same measure qubit to detect errors \cite{kelly_state_2015}. This redefinition has an insignificant impact on code performance, especially when compared to the studied effects of leakage, and so we neglect it from our analysis.

For \strategy{MLR}, we add the multi-level reset (MLR) operation introduced in \cite{mcewen_removing_2021} on the measure qubits at the end of each cycle. Additional pulse shaping on the diabatic return and calibration improvements allow us to achieve gate times of 160~ns without impacting performance.

For \strategy{DQLR}, we first perform the MLR operation on all measure qubits, as in \strategy{MLR}. Following that, we perform the DQLR procedure, which consists of a \textit{LeakageISWAP} gate between pairs of measure and data qubits, and a second reset of the measure qubits. The LeakageISWAP gate is similar to the diabatic CZ gate used in the surface code and bit-flip code cycles, but executes an ISWAP gate in the $\ket{11}-\ket{20}$ subspace. 
We note that this DQLR procedure relies on the high fidelity of the preceding MLR operation; when the measure qubit is prepared in $\ket{0}$, the LeakageISWAP gate removes $\ket{2}$ on the data qubits. Any reset error leaving $\ket{1}$ on the measure qubits will be converted into leakage on the data qubit by the LeakageISWAP gate. Our results show that this error path is sufficiently low in probability so as not to increase the leakage population on the data qubits.

\begin{figure}[tb]
    \centering
    \includegraphics[width=0.48\textwidth]{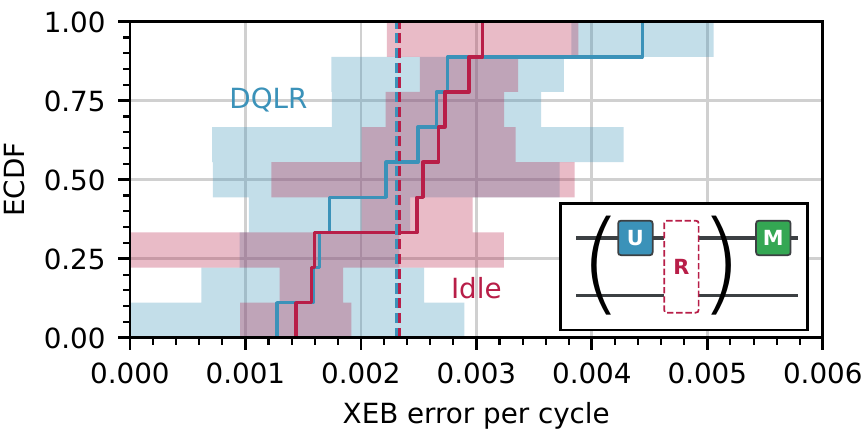}
    \caption[Cross-entropy benchmarking (XEB) of DQLR operation. ]{
    \textbf{Cross-entropy benchmarking (XEB) of DQLR operation. }
        Inferred XEB error per cycle for 9 data qubits when the DQLR operation is applied on the data qubit, compared to XEB error per cycle when the data qubit idles for the equivalent time. Shaded region indicates 1 SD error of inferred XEB error per cycle. Vertical dashed lines indicate the mean XEB error per cycle over all data qubits. (Inset) XEB circuit where random single-qubit unitary rotations (\textbf{U}) are repeatedly applied to the target qubit, interleaved with a reset operation (\textbf{R}), which is either \strategy{DQLR} or idle (\strategy{Idle}). The final state of the target qubit is measured. The cross-entropy between the measured and expected distribution of states is calculated and the resulting XEB error per cycle is inferred. 
    }
    \label{fig:xeb_dqlr}
\end{figure}

\begin{figure*}[t]
    \centering
    \includegraphics[width=\textwidth]{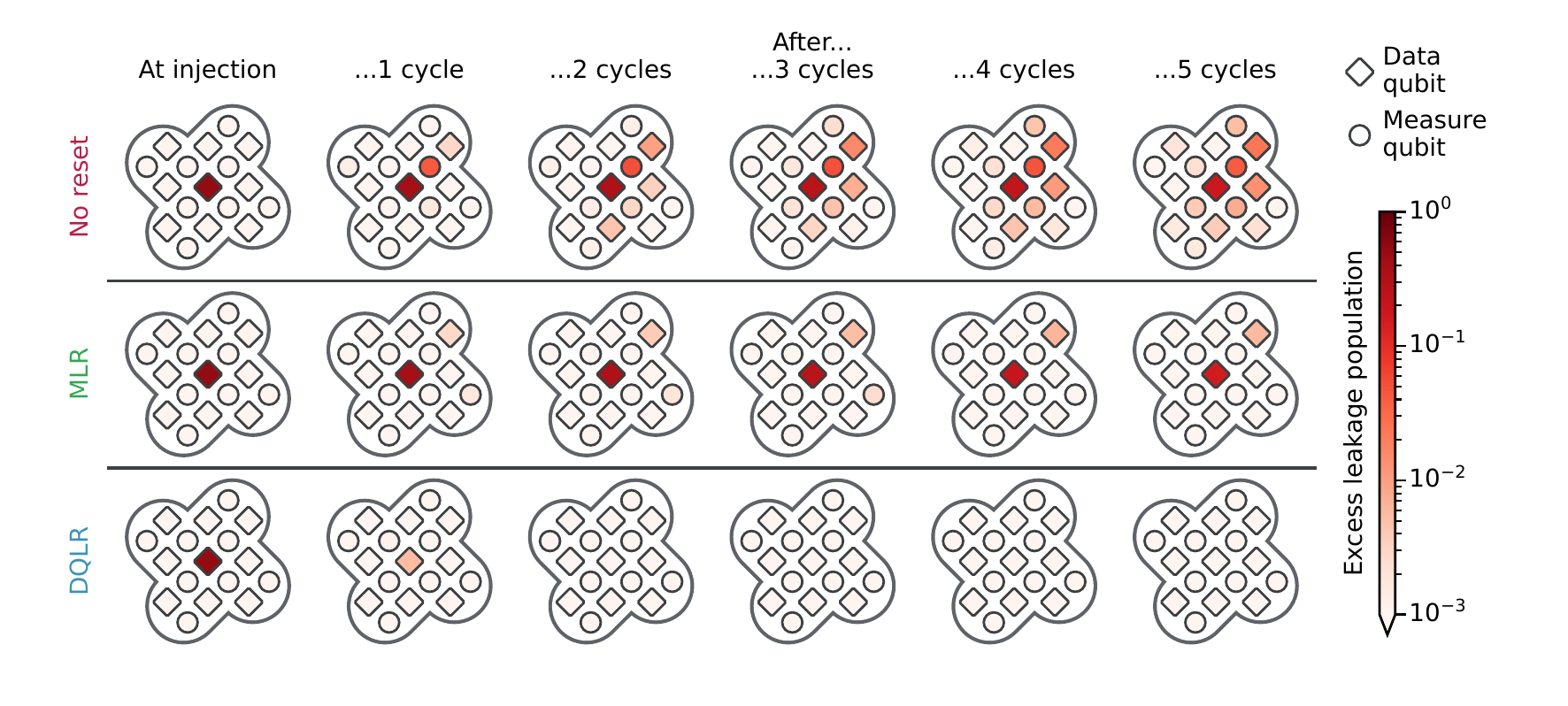}
    \caption[Leakage dynamics in a surface code with different removal strategies. ]{
    \textbf{Leakage dynamics in a surface code with different removal strategies. }
        Comparison of excess leakage population dynamics over 5 cycles for all qubits in a distance-3 surface code after a full $\ket{1}\rightarrow\ket{2}$ leakage injection during the first cycle. With \strategy{No reset}, leakage transport mechanisms lead to increasing leakage population over nearly all qubits involved in the code. With the introduction of \strategy{MLR}, a sink for leakage population is present on measure qubits, mitigating the spread of leakage from leakage transport effects. The central leakage-injected data qubit still remains significantly leaked, even after 5 cycles. Using \strategy{DQLR}, leakage populations on all qubits in the code are brought to about $1\times10^{-3}$ or lower within 2 cycles.
    }
    \label{fig:all_leakage_spread}
\end{figure*}
Ideally, \strategy{DQLR} should not induce additional errors on the data qubit. In particular, when the lower-energy measure qubit is in $\ket{0}$, the LeakageISWAP gate should act as an identity operation on the data qubit computational basis. However, the non-zero time to execute the DQLR procedure introduces incoherent errors caused by relaxation, in addition to coherent errors from miscalibration. We evaluate the impact of the DQLR procedure on the data qubit state using cross-entropy benchmarking (XEB). The inset of \fig{xeb_dqlr} shows the experimental circuit used to evaluate XEB error. The upper and lower qubits mimic the role of data and measure qubits, respectively. The section of the circuit within the parentheses is repeated a variable number of times, and the final state of the upper qubit is measured. The cross-entropy of the measured and expected distributions of states is calculated as a function of repetitions, and then the XEB error per repetition is extracted. In a given repetition, a random unitary \textbf{U} is executed on the upper qubit, followed by a reset operation \textbf{R}. In the case of \strategy{DQLR}, \textbf{R} is substituted with the DQLR procedure. We compare this to the \strategy{Idle} case, where \textbf{R} is replaced with waiting for the duration of the DQLR procedure. We carry this measurement out over the 9 pairs of data and measure qubits corresponding to the pairings used in the distance-3 surface code experiment. By comparing the resulting distribution of XEB error per cycle for \strategy{DQLR} and \strategy{Idle}, we conclude that the DQLR operation does not induce significantly more errors than idling for the equivalent duration. Furthermore, the mean error rate of less than $2.5\times10^{-3}$ per cycle when using the DQLR operation is low enough so that the operation is suitable to be added to a sensitive QEC circuit such as the surface code. We note that leakage is still a relevant consideration in this XEB experiment, and is captured by XEB error per cycle as an incoherent error. Thus, it is possible that the leakage removal properties of \strategy{DQLR} result in underreported XEB error per cycle when compared to \strategy{Idle}, which allows for leakage to accumulate over the course of the XEB circuit.

\section{Effect of reset strategies on leakage dynamics}\label{sup:spread}
As we have demonstrated in the main text, leakage transport can move leakage population from qubit to qubit in a structured circuit such as the surface code. Once a data qubit is leaked, leakage removal strategies must be employed or the leakage population may remain for many QEC cycles and cause additional leakage and leakage-induced error through leakage transport. In \fig{all_leakage_spread}, we evaluate the dynamics of leakage population in a surface code under the three leakage removal strategies discussed in this work. We fully inject leakage at the beginning of the first cycle by performing a $\ket{1}\rightarrow\ket{2}$ rotation on the central data qubit to obtain near-50\% population of $\ket{2}$. We measure excess leakage population, which is defined as the leakage population without injection subtracted from the leakage population with injection. This allows us to separate the contribution of intrinsic heating to leakage population dynamics from the injected leakage population.

For \strategy{No reset}, leakage transport allows for leakage to move freely throughout the qubits surrounding the central data qubit, eventually resulting in measurable excess leakage population in nearly all of the 17 qubits involved in the distance-3 surface code. The lack of leakage removal procedures on either the measure qubits or data qubits leaves $T_1$ energy relaxation of $\ket{2}$ as the primary mechanism for leakage population mitigation. As we showed in Figure 1c of the main text, $T_1$ of $\ket{2}$ can be longer than 10 surface code cycles, and this number is expected to increase as qubit coherence improves and as surface code cycle durations shorten. Hence, relying on energy relaxation is not a viable strategy for leakage removal in QEC. This is partially addressed by \strategy{MLR} by applying a multi-level reset gate on all measure qubits at the end of each cycle.
In \fig{all_leakage_spread}, the effect of this operation appears as reduced excess leakage populations on all measure qubits at the end of every cycle. However, leakage transport within a QEC cycle can still move leakage population beyond nearest-neighbor qubits. This is readily observed in the excess leakage population of the data qubit two sites away from the central data qubit, which exhibits increasing population even though the measure qubit between the two data qubits has its leakage population removed by the MLR operation at the end of every cycle.

\strategy{DQLR} suppresses the ability of leakage to hop between data qubits by directly removing a large fraction of all the data qubits' leakage populations at the end of each cycle. In particular, the central data qubit has its excess leakage population reduced to less than 1\% after the first cycle. Similarly, other data qubits that previously leaked due to leakage transport have their excess leakage populations reduced close to the measurement floor. The shortened lifetime of leakage on all qubits is clearly seen for \strategy{DQLR} after two QEC cycles, where excess leakage population over all qubits is less than $1\times10^{-3}$.

\section{Effect of reset strategies on QEC error detection}\label{sec:qec_error_detection}

\begin{figure}[t]
    \centering
    \includegraphics[width=0.48\textwidth]{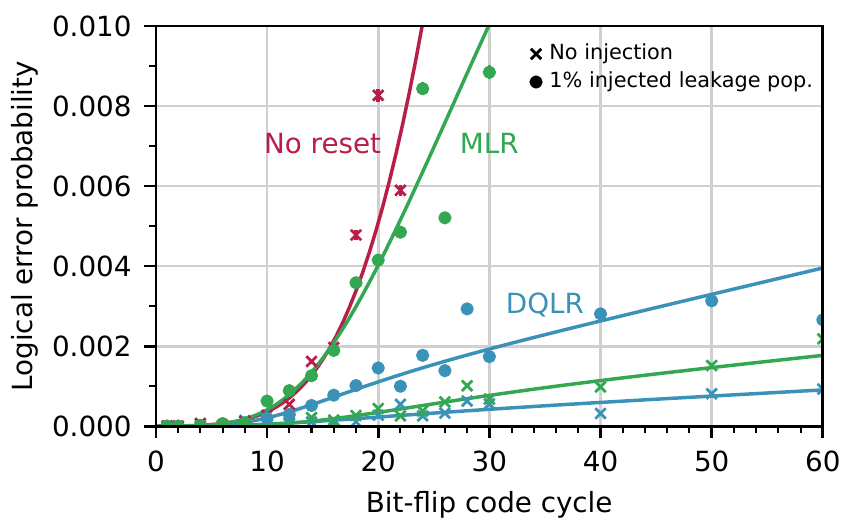}
    \caption[Logical performance of bit-flip code in time. ]{
    \textbf{Logical performance of bit-flip code in time. }
        Logical error probability of the distance-21 bit-flip code over 60 cycles for the three reset strategies \strategy{No reset} (red), \strategy{MLR} (green), and \strategy{DQLR} (blue). For \strategy{MLR} and \strategy{DQLR}, we also execute the code while injecting 1\% leakage population per cycle. Early in the code (fewer than 5 cycles), boundary effects cause all three reset strategies to perform similarly. As the code progresses through more and more cycles, however, logical performance for the three strategies diverge as leakage populations are handled differently.
    }
    \label{fig:bitflip_cycles}
\end{figure}

\begin{figure}[t]
    \centering
    \includegraphics[width=0.48\textwidth]{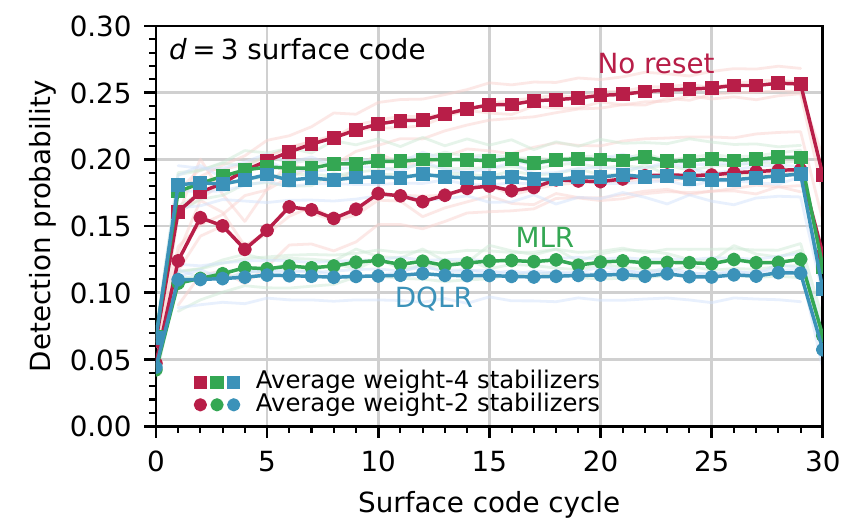}
    \caption[Distance-3 surface code detection probabilities.]{
    \textbf{Distance-3 surface code detection probabilities. }
        Average detection probabilities over 30 surface code cycles for weight-2 (circles) and weight-4 (squares) stabilizers for three leakage removal strategies \strategy{No reset} (red), \strategy{MLR} (green), and \strategy{DQLR} (blue). Individual stabilizer detection probabilities are shown as lighter lines.
    }
    \label{fig:d3_defs}
\end{figure}

In \fig{bitflip_cycles}, we compare the time dynamics of the bit-flip code under the three different reset strategies presented in the main text. We execute a distance-21 bit-flip code over 60 cycles, as described by the circuit in Figure 4d of the main text. For \strategy{MLR} and \strategy{DQLR}, we inject both 0\% and 1\% leakage population in each cycle, whereas for \strategy{No reset} we do not inject leakage. In the first few cycles of code execution, differences in the logical performance of the bit-flip code between the various leakage removal strategies and injection populations are difficult to distinguish -- we attribute this to time-boundary effects where physical errors have not sufficiently accumulated to manifest as a logical error, and statistical limitations where the logical error probability is much smaller than what is resolvable by the number of trials. However, the logical performance for the three leakage removal strategies begins to diverge after about 10 cycles. The accumulation of leakage on measure and data qubits causes a rapid rise of logical error probability for \strategy{No reset}, where it exceeds $1\times10^{-2}$ error by 25 cycles. This is in contrast to \strategy{MLR} and \strategy{DQLR} without leakage injection, which continue to have less than $3\times10^{-3}$ logical error probability through 60 cycles; DQLR has the best performance with about $1\times10^{-3}$ logical error probability at 60 cycles.

Turning to the cases where we inject 1\% leakage population per cycle for \strategy{MLR} and \strategy{DQLR}, we observe a notable distinction in logical error probability scaling over cycle number. With \strategy{MLR}, the logical error probability rises to about $1\times10^{-2}$ by 30 cycles, with a similar qualitative behavior to \strategy{No reset} without leakage injection. However, when we use \strategy{DQLR}, the code sustains logical error probabilities of less than $5\times10^{-3}$ over 60 cycles. This is a signature that QEC scaling in time can be more easily achieved when using DQLR.

In \fig{d3_defs}, we present the average weight-2 stabilizer detection probabilities in addition to the average weight-4 stabilizer values already shown in Figure 5a of the main text. Additionally, we show the detection probabilities associated with the individual stabilizers. We can draw parallel conclusions for the behavior of weight-2 stabilizers as we did for weight-4 stabilizers in the main text. For \strategy{No reset}, the average weight-2 stabilizer detection probabilities rise and do not stabilize over the course of 30 surface code cycles, and even exhibit damped oscillations at early cycles. When using \strategy{MLR}, the detection probability stability improves significantly and the average probability only rises by about $1\times10^{-2}$ over 10 cycles before flattening. However, the best performance and stability is achieved with \strategy{DQLR}, where the average weight-2 stabilizer detection probability remains at 11\% over the course of the entire 30-cycle experiment.

\section{Fitting techniques for logical error probability}\label{sec:fit_leps}

We employ phenomenological models to fit experimental and simulated data. The models are implemented on the logical error per cycle $\varepsilon$ of the QEC code. In experimental and simulated data, we measure logical error probability $p_L$ after $n$ QEC cycles. The conversion between $\varepsilon$ and $p_L$ after $n$ cycles is given by
\begin{align}
    \varepsilon &= \frac{1  - \left(1 - 2p_L\right)^\frac{1}{n}}{2}, \\
    p_L &= \frac{1 - \left(1 - 2\varepsilon\right)^n}{2}.
\end{align}
With respect to injected error population $P$, $\varepsilon$ can be modeled as a power law with an offset,
\begin{align}
    \varepsilon\left(P\right) &= a\left( P + P_{0} \right)^b,
\end{align}
where $a$, $b$, $P_{0}$ are phenomenological free parameters. We do not ascribe a physical meaning to $P_0$ even though it may appear to be ``intrinsic'' error at $P = 0$. We use this power law model to fit the experimental data in Figures 4a and 5b of the main text, as well as the numerical simulations in Figure 4a of the main text.

In order to model $1/\Lambda_{5/7}$ as a function of injected leakage population $P_L$ in Figure 5c of the main text, we take the inverse of the ratio between distance-5 and distance-7 logical error per cycle,
\begin{align}
    \Lambda_{5/7}\left(P_L\right) &= \frac{\varepsilon_5\left(P_L\right)}{\varepsilon_7\left(P_L\right)}. \label{eqn:lambda}
\end{align}

For logical error probability $p_L$ with respect to cycle $n$ of a bit-flip code operated well below threshold and in the presence of leakage dynamics (\fig{bitflip_cycles}), we use a Gompertz model to describe logical error rate $\varepsilon$,
\begin{align}
    \varepsilon\left(n\right) &= a \exp\left(-b \exp\left(-cn\right)\right),
\end{align}
where $a$, $b$, and $c$ are phenomenological free parameters. A Gompertz model can partly capture the transient dynamics of $\varepsilon$ at small $n$, where time-boundary effects and still-increasing leakage populations make $\varepsilon$ highly dependent on $n$. At large $n$, $\varepsilon$ tends to a constant as leakage populations stabilize.

\section{Surface code simulations well below threshold}\label{sec:sims}

\begin{figure*}[t]
    \centering
    \includegraphics[width=\textwidth]{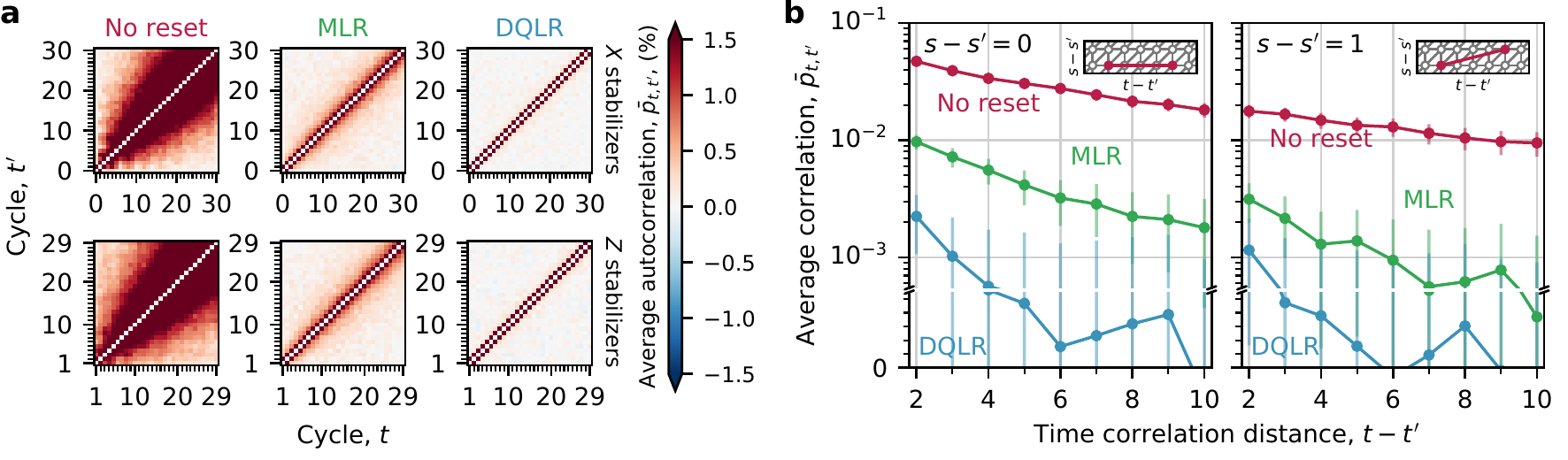}
    \caption[Surface code $p_{ij}$ correlations. ]{
        \textbf{Surface code $p_{ij}$ correlations. }
        a) Average autocorrelated $p_{ij}$ matrices $\bar{p}_{t,t'}$ of $X$ and $Z$ stabilizers under different leakage removal strategies. Correlations along the upper and lower diagonals ($p_{t,t'}$ where $|t - t'| = 1$) represent independent timelike errors, whereas non-local correlations ($p_{t,t'}$ where $|t - t'| > 1$) can be primarily attributed to leakage-induced correlated errors. For \strategy{No reset}, non-local correlations intensify as the code is executed in time, suggesting increasing leakage-induced correlated errors from growing leakage population. With \strategy{MLR}, non-local correlations are reduced but remain, which are manifestations of data qubit leakage-induced correlated errors. For \strategy{DQLR}, complete leakage removal over all qubits results in suppression of non-local correlations to about $1\times10^{-3}$ or lower.
        b) (Left) Averaged autocorrelation $\bar{p}_{t,t'}; s-s'=0$ over all stabilizers for cycles $t$ and $t'$ in 19--29, inclusive, under different leakage removal strategies. Average non-local correlation magnitudes do not exceed $2\times10^{-3}$ for \strategy{DQLR}, whereas \strategy{No reset} and \strategy{MLR} have exponential decays in correlation magnitude with respect to correlation distance. The inset shows an example of a distance-4 long-time error edge on a detection graph.
        (Right) Averaged nearest-neighbor correlation $\bar{p}_{t,t'}; s-s'=1$ over all stabilizers, extracted under the same conditions as the averaged autocorrelation values. The inset shows an example of a distance-4 long-diagonal error edge on a detection graph.
        }
    \label{fig:timelike_pij}
\end{figure*}

To gain insight into the future importance of leakage removal for scaling quantum error correction, we performed numerical simulations of distance-5 and distance-7 surface codes and evaluated their logical performance subject to different levels of leakage injection carried out using the circuit shown in the inset of Figure 5b of the main text. We performed these simulations using a Kraus operator simulation detailed in \cite{google_quantum_ai_suppressing_2022}. We include operators that accurately reflect leakage transport, leakage phase errors, and the leakage removal parameters for both the \strategy{MLR} and \strategy{DQLR} strategies, but do not include any sources of leakage in the baseline error model at zero leakage injection.
The error model for the baseline simulations is detailed in \tbl{noise_model}.
We repeat these simulations with varying amounts of injected leakage population under two leakage removal strategies, \strategy{MLR} and \strategy{DQLR}, which is presented in the Figure 5c of the main text.

We fit the data from the numerical simulations in Figure 5c of the main text with a power law ratio model (\eqn{lambda}). Additionally, we fit the data for \strategy{DQLR} with a line. The high degree of linearity ($R^2 = 0.983$, $1/\Lambda_{5/7} \approx 111 \times P_L + 0.2$) suggests that the surface code error budget $1/\Lambda_{5/7}$ for leakage under \strategy{DQLR} can be linearized and modeled as an uncorrelated error source.

\begin{table}[b]
    \caption[Hypothetical device error model for simulations carried out in Figure 5c of the main text.]{
    \textbf{Hypothetical device error model for simulations carried out in Figure 5c of the main text.}
    }
    \begin{ruledtabular}
    \begin{tabular}{l c}
    \textrm{Parameter}&
    \textrm{Value}\\
    \colrule
    Single qubit gate Pauli error & $2\times10^{-4}$\\
    CZ gate Pauli error & $1\times10^{-3}$\\
    Readout and reset error & $1\times10^{-2}$ \\
    Idling Pauli error from relaxation & $3\times10^{-3}$\\
    \;\;\;\; from dynamical decoupling & $1\times10^{-3}$ \\
    Qubit $T_1$ & 75 $\mu$s \\
    Qubit $T_2$ & 75 $\mu$s \\
    Single qubit gate time & 15 ns \\
    CZ gate time & 25 ns \\
    Combined readout and reset time & 300 ns \\
    \end{tabular}
    \end{ruledtabular}
    \label{tbl:noise_model}
\end{table}

\section{Error correlations in the surface code experiment}\label{sec:pij}

One technique to evaluate error correlations in the surface code is to employ $p_{ij}$ correlation matrices \cite{obrien_density-matrix_2017, mcewen_removing_2021, google_quantum_ai_exponential_2021, google_quantum_ai_suppressing_2022}. By analyzing the data presented in \fig{d3_defs} with $p_{ij}$ correlation matrices, we can elucidate the presence of error correlations in space and time over the course of the QEC experiment.

For $p_{ij}$ matrices, $i$ and $j$ correspond to nodes in the detection graph, each of which have stabilizer (or \textit{space}, $s$) and time ($t$) coordinates; $i = \left(s, t\right)$, $j = \left(s', t'\right)$. To first focus on correlations in time, we average the matrix elements $p_{ij}$ with $s = s'$ over all $t$ and $t'$, producing autocorrelation matrices $\bar{p}_{t,t'}$, defined as
\begin{align}
    \bar{p}_{t,t'} &= \frac{\sum\limits_{s = s'}{p_{ij}}}{\sum\limits_{s = s'} 1}.
\end{align}
The autocorrelation matrix $\bar{p}_{t,t'}$ reports the average of the probabilities of error graph edges between nodes $i$ and $j$ associated to the same stabilizer $s$ and for arbitrary time separation $|t - t'|$.

In \fig{timelike_pij}a, we show averaged autocorrelation matrices $\bar{p}_{t,t'}$ for both $X$- and $Z$-basis stabilizers over the 30 cycle experiment under the three leakage removal strategies investigated in this work. $Z$-basis stabilizers do not report detection probabilities at the time-boundary cycles 0 and 30 and thus do not have $\bar{p}_{t,t'}$ elements associated with those cycles. Independent Pauli errors present themselves as detection events on consecutive cycles $|t - t'| = 1$. In such an ideal setting with only fully independent errors, we expect non-local correlations to vanish, \textit{i.e.} $\bar{p}_{t,t'} = 0$ where $|t - t'| > 1$. Leakage and other time-non-local error sources will break this assumption and force those $\bar{p}_{t,t'}$ elements to be non-zero.

For \strategy{No reset}, non-local correlations immediately appear at cycle 1 and increase in intensity over the course of the 30-cycle experiment. The high values for non-local $\bar{p}_{t,t'}$ matrix elements greater than 1\% at large correlation distances $|t - t'|$ indicate that non-local correlations are significant contributors to logical performance degradation, and that QEC cannot be practically scaled in time under these conditions \cite{fowler_coping_2013}.

For \strategy{MLR}, non-local correlations are visibly reduced when compared to \strategy{No reset}. Still, an impactful degree of non-local correlation remains, primarily stemming from data qubit leakage and resultant leakage-induced correlated errors, such as CZ gate phase shifts highlighted in Figure 2e of the main text.

For \strategy{DQLR}, nearly all non-local correlations are heavily suppressed in the experiment. This is seen by very small correlation magnitudes less than 0.2\% at correlation distances greater than 1. Qualitatively, the effective suppression of time-non-local correlations suggests we are much closer to fulfilling the QEC requirement of uncorrelated errors. Furthermore, this is evidence that our DQLR procedure does not introduce additional unwanted correlations within the experiment.

We isolate and average $\bar{p}_{t,t'}$ matrix elements for cycles $t$ and $t'$ in 19--29, inclusive, in \fig{timelike_pij}b. This offers a quantitative profile to the degree of non-local correlations present in the surface code as a function of time correlation distance $t - t'$, which correspond to the probability of encountering \emph{long-time error edges} in the detection graph. Again, under ideal circumstances with fully independent errors, the correlation strength $\bar{p}_{t,t'}$ should be 0 at all $t - t'$ greater than 1. \strategy{DQLR} most closely approximates this condition, where $|\bar{p}_{t,t'}|$ exceeds $2\times10^{-3}$ only for distance-2 correlation and otherwise is below 0.2\% for all distances up through 10. The 1 SD error bars for \strategy{DQLR} suggest that we cannot resolve variations in $|\bar{p}_{t,t'}|<1\times10^{-3}$ for these non-local correlations. However, for \strategy{MLR}, $|\bar{p}_{t,t'}|$ is about an order of magnitude higher at 1\% for distance-2 correlation, and slowly decays with distance, remaining above 0.1\% even at distance-10. Finally, \strategy{No reset} never has $\bar{p}_{t,t'}<1\%$ for any of the correlation distances considered here.

In order to evaluate correlations across space, we can perform a similar analysis but now with nearest-neighbor stabilizers,
\begin{align}
    p_{t,t'} &= \frac{\sum\limits_{s - s' = 1}{p_{ij}}}{\sum\limits_{s - s' = 1} 1}.
\end{align}
The resulting average time correlations, which indicate the probability of encountering a \emph{long-diagonal error edge}, are shown in the right panel of \fig{timelike_pij}b. By only considering time correlation distances of $t - t' > 1$, we primarily exclude contributions from CZ gate errors, which manifest as \emph{diagonal edges} at $s - s' = 1$ and $t - t' = 1$ \cite{google_quantum_ai_suppressing_2022}. The remaining correlations at $t - t' > 1$ then predominantly arise from leakage and crosstalk and should be zero under ideal circumstances without these error sources.

Similar to the average correlation $\bar{p}_{t,t'}$ for $s - s' = 0$, we observe the same hierarchy of correlation strengths for $s - s' = 1$ under \strategy{No reset}, \strategy{MLR}, and \strategy{DQLR}. \strategy{No reset} performs the worst with correlation strengths at about 1\% for all correlation distances studied here. \strategy{MLR} performs considerably better, but still has measurable correlations at time distances as large as 6 cycles. Given that we desire zero correlation at all time distances, \strategy{DQLR} most closely approximates this condition, with $\left|\bar{p}_{t,t'}\right|$ at 0.1\% for distance-2 correlations, and otherwise well below 0.1\% for longer time distances.

These observed correlation strengths suggest that if scalable QEC requires near-independent errors, complete leakage removal on all qubits must be carried out in some form. The \strategy{DQLR} strategy presented in this work provides one pathway to reaching that requirement. As a corollary, we show that \strategy{No reset} and \strategy{MLR} cannot support this requirement under existing leakage rates and transport mechanisms.

\end{document}